\documentstyle[11pt]{article}

\begin{document}
\newcommand{\be}{\begin{equation}}
\newcommand{\ee}{\end{equation}}
\newcommand{\bea}{\begin{eqnarray}}
\newcommand{\eea}{\end{eqnarray}}
\newcommand{\bref}[1]{(\ref{#1})}
\newcommand{\ga}{\alpha}
\newcommand{\gb}{\beta}
\renewcommand{\th}{\theta}
\newcommand{\dr}{\partial_r}
\newcommand{\dth}{\partial_\th}
\newcommand{\Lagf}{{\cal L}_{\rm f}}

\newcommand{\Mab}{{M_{\ga \gb}}}
\newcommand{\Cab}{{C_{\ga \gb}}}
\newcommand{\nfer}{{n_{\rm f}}}
\newcommand{\np}{{n_+}}
\newcommand{\nm}{{n_-}}
\newcommand{\nh}{{n}}
\newcommand{\nr}{N_R}
\newcommand{\nl}{N_L}
\newcommand{\TF}[1]{{\rm I}\left[{#1}\right]}
\newcommand{\pos}[1]{\left[{#1}\right]_+}
\newcommand{\Iall}{{\cal I}}
\newcommand{\Idiff}{{\cal I}_\Delta}

\newcommand{\smat}[4]{\left(\begin{array}{cc} {#1} & {#2} \\ 
				{#3} & {#4} \end{array} \right)}
\newcommand{\phnb}[1]{\Phi_{{#1}}}
\newcommand{\ph}[1]{\mbox{$\phnb{{#1}}$}}
\newcommand{\pharrow}[1]{\stackrel
	{ \mbox{\scriptsize $\phnb{{#1}}$} }{\longrightarrow}}

\title{\bf Cosmic String Zero Modes and Multiple Phase Transitions}

\author{Stephen C. Davis$^{1}$\footnote{S.C.Davis@damtp.cam.ac.uk}, 
Anne-Christine Davis$^{1}$\footnote{A.C.Davis@damtp.cam.ac.uk} and
Warren B. Perkins$^{2}$\footnote{w.perkins@swansea.ac.uk} \\ \\
\em ~$^1$Department of Applied Mathematics and Theoretical Physics,\\
\em University of Cambridge, Cambridge, CB3 9EW, UK. \\ \\
\em ~$^2$Department of Physics, University of Wales Swansea,\\
\em  Singleton Park, Swansea, SA2 8PP, Wales}

\maketitle

\begin{abstract}

The zero modes and current carrying capability of a cosmic string
formed at one phase transition can be modified at subsequent phase
transitions. A new, generalised index theorem is derived that is
applicable to theories with multiple phase transitions. This enables
us to investigate the fate of string zero modes during sequences of phase
transitions in a variety of models. Depending on the couplings that
the breaking introduces, the zero modes may be destroyed and the
superconductivity of the string removed, and thus vortons dissipate.
We discuss the features of the theory that are required to produce this 
behaviour and consider the implications of spectral flow.

\end{abstract}

\vfill

\begin{flushright} DAMTP/97-39 \\ SWAT/140 \end{flushright}

\vfill

\newpage

\section{Introduction}

Topological defects, and in particular cosmic strings, arise in many
grand unified theories \cite{string book}. They are likely to be copiously 
produced during phase transitions in the early universe. A network of 
cosmic strings could explain the observed anisotropy in the microwave
background radiation and the large scale structure of the Universe
\cite{string book}. In the past few years it has been realised that
they may have considerably richer microstructure than previously
realised \cite{wbp&acd}. In particular, the presence of fermion zero
modes \cite{Jackiw} in the spectrum of a cosmic string has profound
implications for the cosmology of the defects. Zero modes render the
string superconducting \cite{witten}, allowing currents of up to
$10^{20}$ Amps to flow along the string. Stable relics, vortons, can
form as collapsing string loops are stabilised by the angular momentum
of the trapped charge carriers \cite{vorton}. These relics can
overclose the Universe unless stringent constraints are applied to the
string model \cite{vortbounds}.

The cosmological implications of the vortons are most pronounced when
the Universe has become matter dominated. If they decay during the
era of radiation domination the cosmological catastrophe may be
avoided and the vorton bounds evaded. However, it has recently been
realised that subsequent phase transitions can have a considerable effect
on the microphysics of cosmic strings. Zero modes on cosmic strings
can be both created \cite{acd&wbp} and destroyed \cite{SO10strings},
thus creating or destroying currents on strings. In this paper we discuss 
the fate of vortons and string superconductivity as the strings encounter
subsequent phase transitions in a systematic fashion.

In section \ref{Index} we derive an index theorem giving the number of
zero modes for a general mass matrix. Whilst index theorems have been
derived before \cite{{Weinberg},{Ganoulis}} they have been more restrictive 
in their validity and have only been able to determine the difference
in right-moving and left-moving zero modes. Our index theorem has much
more general applicability and can give a bound on the number of zero modes. 

In section \ref{SO10} we discuss the fate of zero modes on strings formed at
the breaking of an SO(10) Grand Unified symmetry \cite{SO10strings}. At the 
electroweak phase transition these zero modes acquire a small mass which 
leads to dissipation of the string current. This allows
vortons to decay and weakens the cosmological bounds on such models
\cite{acd&wbp3}.

In other models the zero modes survive subsequent transitions allowing
the associated vortons to persist. Such behaviour is displayed by a
toy model discussed in section \ref{Persistent}, where we explicitly
construct the zero mode solutions of the Dirac equation after the
symmetry breaking.

We compare these two models in section \ref{Spectral} and consider the
implications of spectral flow. An important feature
that allows zero modes to be removed is the presence of a
particle that mixes with the zero mode after the transition. The
implications of such couplings for current build up before the
transition are also considered. Finally, we summarise our conclusions.

\section{Zero Mode Index}
\label{Index}
\newcommand{\half}{\mbox{$\frac{1}{2}$}}

Cosmic strings form in models with vacuum manifolds which are not
simply connected. For example in a $U(1)$ model, with potential
$(|\phi|^2 - \eta^2)^2$, stable solutions exist with $\phi = \eta e^{i\th}$
at $r=\infty$. In order for the total energy to be finite, a non-zero
gauge field is needed to give a vanishing covariant derivative at
$r=\infty$. In a more general theory, involving a larger group, $G$, string
solutions take the form
\bea
\phi(r,\th) = e^{iT_s\th} \phi(r) \ , \ 
A_\th = \frac{1}{er} T(r) \ ,
\eea
where $T_s$ is a generator of $G$ that is broken by $\phi$. The choice
of $T_s$ is restricted by the fact that $\phi$ must be single valued. $\phi(r)$
is equal to the usual VEV of $\phi$ at $r=\infty$, and must be regular
at $r=0$. $T(r)$ obeys $T(0)=0$, $T(\infty)=T_s$.

In a general theory, $T_s$ will affect different components of $\phi$
differently. This means that the various parts of $\phi$ can have a
wide range of winding numbers. In a theory with multiple phase
transitions, the additional Higgs fields will be affected in the same
way. It may also be necessary to alter $T_s$ at phase transitions to
make the new Higgs fields single valued.

In a theory with $\nfer$ two-component fermions, the fermionic part of
the Lagrangian is 
\be 
\Lagf = \bar{\psi}_\ga i\sigma^\mu D_\mu \psi_\ga 
- \frac{1}{2}i\bar{\psi}_\ga \Mab \psi^c_\gb + (\mbox{h.c.})
\label{Lag}
\ee
where $\psi^c_\gb = i\sigma^2\bar{\psi}_\gb^T$. If $\Mab$ depends on
$\th$, as would be expected if $\Mab$ arose from the Higgs field of
the string, then it is possible that the field equations will have
non-trivial zero energy solutions. Solutions with only $r$ and $\th$
dependence can be split up into eigenstates of $\sigma^3$:
$\psi^L_\ga$, $\psi^R_\ga$. Such solutions have zero energy. If we
solve the equations of motion in the background of a cosmic string,
the field equations become
\be
e^{i\th}\left(\dr + \frac{i}{r}\dth + eA_\th \right) \psi^L_\ga 
+ M_{\ga \gb} \psi^{L\ast}_\gb = 0 \ ,
\label{FEleft}
\ee
\be
e^{-i\th}\left(\dr - \frac{i}{r}\dth - eA_\th \right) \psi^R_\ga 
- M_{\ga \gb} \psi^{R\ast}_\gb = 0 \ ,
\label{FEright}
\ee
where $A_\th$ is the string gauge field.
If $z$ and $t$ dependence is added to the solutions they will
correspond to currents flowing along the string. Their direction is
left for those corresponding to \bref{FEleft}, and right for
\bref{FEright}. In order to be physically relevant the solutions must
be normalisable. Let $\nl$ and $\nr$ be the number of such solutions
to \bref{FEleft} and \bref{FEright} respectively. We attempt to derive
an expression for them by generalising the analysis in ref.\
\cite{Jackiw}, which involves removing the $\th$ dependence of the
problem, and then considering solutions near $r=\infty$ and $r=0$.

Choose the $\psi_\ga$s to be eigenstates of the string gauge field,
with eigenvalues $q_\ga$. The $q_\ga$ will depend on the fermion
charges and the winding numbers of the various components of the Higgs
fields. Since the mass terms in \bref{Lag} are gauge invariant, the
angular dependence of the mass matrix must be 
\be
M_{\ga \gb}(r,\th) = \Cab(r)e^{i(q_\ga + q_\gb)\th} \ 
\ \ \ (\mbox{no summation}) \ .
\label{MassNth}
\ee
The $\th$ dependence can also be factored out of the $\psi_\ga$s. 
\bea
\psi^L_\ga &=& e^{i(q_\ga - \frac{1}{2})}
		(U^L_\ga(r)e^{il\th} + V^{L\ast}_\ga(r)e^{-il\th}) \ ,\\
\psi^R_\ga &=& e^{i(q_\ga + \frac{1}{2})}
		(U^R_\ga(r)e^{il\th} + V^{R\ast}_\ga(r)e^{-il\th}) \ . 
\label{PsiNth}
\eea

First consider left moving zero modes.
Putting (\ref{MassNth},\ref{PsiNth}) into \bref{FEleft} gives
equations for $U^L_\ga$ and $V^L_\ga$.
As $r \rightarrow \infty$, $\Cab = O(1)$ and $eA_\th =
O(1/r)$, so
\bea
\dr U^L_\ga + \Cab(\infty) V^L_\gb = 0 \ , \label{BigrFEU} \\
\dr V^L_\ga + \Cab(\infty) U^L_\gb = 0 \ . \label{BigrFEV}
\eea
Diagonalising $\Cab$ gives $2\nfer$ complex solutions
proportional to $\exp(\pm \lambda_\ga)$, where $\lambda_\ga$ are $\Cab$'s
eigenvalues. Thus, assuming all $\lambda_\ga \neq 0$, exactly $\nfer$
of these large $r$ solutions are normalisable at $r=\infty$.

The Higgs fields, and hence $\Mab$, are regular at the origin, so
as $r \rightarrow 0$, $\Cab = O(1)$, and $eA_\th = O(r)$. Thus
\bea
\left(\dr - \frac{q_\ga - \frac{1}{2} + l}{r}\right) U^L_\ga 
&+& \Cab V^{L\ast}_\gb = 0 \ , \label{SmallrFEU} \\
\left(\dr - \frac{q_\ga - \frac{1}{2} - l}{r}\right) V^L_\ga 
&+& \Cab U^{L\ast}_\gb = 0 \ . \label{SmallrFEV}
\eea 
To leading order, the small $r$ solutions are
\be \begin{array}{lll}
U^L_\ga &\sim& r^{q_\ga - \frac{1}{2} + l} \ , \\
V^L_\gb &\sim& O(1)r^{q_\gb + \frac{1}{2} + l} \ \ \forall \gb \ , \\
U^L_\gb &\sim& O(1)r^{q_\gb + \frac{3}{2} + l} \ \ \forall \gb \neq \ga \ ,
\end{array} \label{leftUsol}
\ee
where each choice of $\ga=1 \ldots \nfer$ gives one complex solution, and
\be \begin{array}{lll}
V^L_\ga &\sim& r^{q_\ga - \frac{1}{2} - l} \ , \\
U^L_\gb &\sim& O(1)r^{q_\gb + \frac{1}{2} - l} \ \ \forall \gb \ , \\
V^L_\gb &\sim& O(1)r^{q_\gb + \frac{3}{2} - l} \ \ \forall \gb \neq \ga \ ,
\end{array} \label{leftVsol}
\ee
which gives a total of $2\nfer$ independent complex solutions. For
given $l$ and $\ga$, \bref{leftUsol} will be normalisable (for small $r$) if 
$l \geq -q_\ga + 1/2$. If $l \leq q_\ga - 1/2$ then
\bref{leftVsol} will be normalisable. Thus for a given $l$ the number
of well behaved small $r$ solutions is
\be
\nl^0 (l) = \sum^\nfer_{\ga=1} \TF{l \leq q_\ga - \half} + 
	\TF{l \geq -q_\ga + \half} 
\ee
where $\TF{X}$ equals 1 if $X$ is true, and 0 if $X$ is false.
What we are actually interested in is the number of solutions that are
normalisable for all $r$ ($\nl(l)$). Each such solution will be equal to
some combination of the $\nfer$ well behaved solutions to
(\ref{BigrFEU},\ref{BigrFEV}) at large $r$, and a combination of the $\nl^0(l)$
suitable solutions to (\ref{SmallrFEU},\ref{SmallrFEV}) for small $r$.
If there are only $\nfer$, or less, suitable small $r$ solutions, then
in general any combination of the large $r$ solutions will not be well
behaved at $r=0$. If there are $\nfer + k$ suitable small $r$
solutions, then $k$ independent combinations of the large $r$
solutions will be well behaved everywhere. It may be possible to get
more solutions by fine tuning the theory, in which case the index
derived would be a lower bound.

The number of normalisable solutions for a given $l$ is
\be
\nl(l) = \pos{\nl^0 (l) - \nfer} \ ,
\ee
where $\pos{x}$ is defined to be equal to zero if $x<0$, and $x$ if $x \geq 0$.

This is not true if the equations obtained from \bref{FEleft} can be
split into several independent sets. This will occur when 
$\Mab$ is a direct product of mass matrices. In this case the
mass matrix can be split up into smaller matrices, which can be
analysed individually. Even when $\Mab$ is not a direct product
of other matrices, it may still be possible to split the equations
into two independent sets. This case will be considered separately later. 

Since $U^L_\ga$ and $V^L_\ga$ are determined by real equations, each
complex solution gives two real solutions. This suggests that the total
number of left moving zero modes, $\nl$, is $2\sum_l \nl(l)$. However,
as can be seen from \bref{PsiNth}, solutions for $l=k$ and $l=-k$ are
equal. For $l=0$ $U^L_\ga = \pm V^L_\ga$, so one of $\psi^L_\ga$'s
solutions is zero. Thus the total number of independent real solutions
is
\be
\nl = \sum_l \nl(l) = \sum_l \pos{\sum^\nfer_{\ga=1} \left(\TF{l \leq q_\ga 
	- \half} + \TF{l \geq -q_\ga + \half}\right) - \nfer} \ .
\label{nl1}
\ee
The summation is over all values of $l$ that give single valued
$\psi$. Since all the Higgs fields which make up $M_{\ga \gb}$ are
single valued, \bref{MassNth} implies all $q_\ga$ or all $q_\ga - 1/2$ are
integers (assuming $M_{\ga \gb}$ is not a product of smaller
matrices), in which case respectively $l - 1/2$ or $l$ is an integer.

A similar analysis can be applied to right moving zero modes. For large
$r$ the behaviour is the same. For small $r$, solutions are well
behaved if $l \geq q_\ga + 1/2$ or $l \leq -q_\ga - 1/2$. This gives
\be
\nr = \sum_l \nr(l) = \sum_l \pos{\sum^\nfer_{\ga=1} \left(\TF{l \leq 
	-q_\ga - \half} + \TF{l \geq q_\ga + \half}\right) - \nfer} \ .
\label{nr1}
\ee

If $q_\ga$ is positive then one or two of the $\TF{\ldots}$ terms will
be non-zero. If $q_\ga$ is negative, one or zero of them will be non-zero.
By splitting $q_\ga$ into positive and negative eigenvalues, and ordering
them, \bref{nl1} and \bref{nr1} can be simplified. If there are $\np$
positive and $\nm$ negative $q_\ga$s then, after
reordering, $q_\ga$ = ($p_1$, $p_2$ $\ldots$ $p_\np$, $-g_1$, $-g_2$
$\ldots$ $-g_\nm$, 0 $\ldots$ 0), where $p_j,g_j > 0$. Clearly if the
string gauge eigenvalues are not integer, there are no zeros. The
$\TF{\ldots}$ terms in \bref{nl1} and \bref{nr1} can be combined to give
\be
\nl = \sum_l \pos{
\sum^\np_{j=1} \TF{-p_j + \half \leq l \leq p_j - \half}
- \sum^\nm_{j=1} \TF{-g_j - \half < l < g_j + \half}   } \ ,
\label{nl2} \\
\ee \be
\nr = \sum_l \pos{
\sum^\nm_{j=1} \TF{-g_j + \half  \leq l \leq g_j - \half}
- \sum^\np_{j=1} \TF{-p_j - \half < l < p_j + \half}  } \ .
\label{nr2}
\ee
These expressions can be further simplified if the eigenvalues are
ordered. If $p_1 \geq p_2 \geq \ldots \geq p_\np > 0$ and 
$g_1 \geq g_2 \geq \ldots \geq g_\nm > 0$, it is possible to evaluate
the $l$ summation by considering cancellation of the $p_j$ and $g_j$ terms. 
This gives
\bea
\nl &=& \sum^{\min(\nm,\np)}_{j=1} 2\pos{p_j - g_j}
			+ \sum^\np_{j=\nm+1} 2 p_j \ , \label{nl3} \\
\nr &=& \sum^{\min(\nm,\np)}_{j=1} 2\pos{g_j - p_j} 
			+ \sum^\nm_{j=\np+1} 2 g_j \ . \label{nr3}
\eea
Taking the difference of these results gives
\bea
\Idiff = \nl - \nr &=& \sum^{\min(\nm,\np)}_{j=1} 2(p_j - g_j) +
\sum^\np_{j=\nm+1} 2p_j - \sum^\nm_{j=\np+1} 2g_j \label{diffindex}\\
 &=& \sum^\nfer_{\ga=1} 2q_\ga 
 = \frac{1}{2 \pi i} \left[\ln \det M \right]^{2\pi}_{\th=0} \ .
\eea
This is in agreement with another index theorem obtained elsewhere
\cite{Weinberg,Ganoulis}. The other index theorem was obtained by a
different method and only gave $\Idiff$, not $\nl$ and $\nr$.

The total number of zero modes is
\be
\Iall = \nl+\nr = \sum^{\min(\nm,\np)}_{j=1} 2\left|p_j - g_j\right| +
\sum^\np_{j=\nm+1} 2p_j + \sum^\nm_{j=\np+1} 2g_j \ ,
\label{allindex}
\ee
where only one of the last 2 terms contributes, depending on whether
$\np$ or $\nm$ is bigger. This is also true of \bref{diffindex} and
(\ref{nl3},\ref{nr3}).

If $\Iall$ is to be zero, then for every positive $q_\ga$ there
must be one negative $q_\gb$ with the same magnitude. If every fermion
field couples to a Higgs field with winding number zero, this will be
the case.

The above approach fails if $\Cab$ is of the form
\be
\smat{0}{J_{\ga \gb}}{K_{\ga \gb}}{0} \ .
\ee
As $\Cab$ is assumed to have no zero eigenvalues, $J_{\ga \gb}$
and $K_{\ga \gb}$ are both $\nh \times \nh$ matrices, where $\nh = \nfer/2$.
In this case when (\ref{MassNth},\ref{PsiNth}) are substituted into
\bref{FEleft}, two independent sets of equations are
obtained. Expressions for $\nl$ and $\nr$ can found by considering
just one set of these solutions. Putting
\be
\psi^L_\ga = e^{i(q_\ga - \frac{1}{2})} 
\left\{ \begin{array}{ll}
	U^L_\ga(r)e^{il\th} & \ga = 1 \ldots \nh \\
	V^{L\ast}_\ga(r)e^{-il\th}) & \ga = \nh +1 \ldots \nfer
\end{array} \right. 
\ee
and \bref{MassNth} into \bref{FEleft} gives
(\ref{BigrFEU},\ref{BigrFEV}) for large $r$ and
(\ref{SmallrFEU},\ref{SmallrFEV}) for small $r$, but with a more restricted
range on the indices. For large $r$, $\nh$ of the $\nfer$ complex
solutions are normalisable. For a given $l$ at small $r$, there is one
normalisable solution for each $q_\ga$ satisfying 
$l \geq -q_\ga + 1/2$ ($\ga = 1 \ldots \nh$) or 
$l \leq q_\ga - 1/2$ ($\ga = \nh +1 \ldots \nfer$).

Matching the solutions for large and small $r$ gives
\be
\nl(l) = \pos{\sum^\nh_{\ga=1} \left( \TF{l \leq q_{\ga+\nh} - \half} 
		+ \TF{l \geq -q_\ga + \half}\right) - \nh} \ . 
\label{nlspec}
\ee
The solutions for different $l$ are independent (unlike the previously
considered cases), so the total number of real solutions ($\nl$) is just twice
the number of complex solutions, thus $\nl = 2\sum_l \nl(l)$. 
A similar expression can be obtained for $\nr$.

The $\TF{\ldots}$ terms in \bref{nlspec} can be combined. Defining 
$g_j = -q_j$ and $p_j = q_{j+\nh}$ for $j = 1 \ldots \nh$,
\bref{nlspec} and the corresponding expression for $\nr$, become 

\bea
\nl &=& 2\sum_l \pos{\sum^\nh_{j=1} \left( 
\TF{g_j + \half \leq l \leq p_j - \half} - \TF{p_j - \half < l < g_j + \half}
\right)} \ , \\
\nr &=& 2\sum_l \pos{\sum^\nh_{j=1} \left( 
\TF{p_j + \half \leq l \leq g_j - \half} - \TF{g_j - \half < l < p_j + \half}
\right)} \ .
\eea
If the string gauge eigenvalues are reordered, so that $p_1 \geq
\ldots \geq p_{\nh}$ and $g_1 \geq \ldots \geq g_{\nh}$, the
expressions reduce to \bref{nl3} and \bref{nr3}, with
$\np=\nm=\nh$. This is not identical to the previous result, since
$p_j$ and $g_j$ are defined differently.

When there are just two fermion fields involved, all the results reduce to 
\bea
\Idiff &=& 2(q_1 + q_2) \ , \\
\Iall &=& |\Idiff| \ .
\label{n2fer}
\eea

\section{An $SO(10)$ GUT with Strings}
\label{SO10}
\newcommand{\mgut}{m_{\rm GUT}}
\newcommand{\mup}{m_{\rm u}}

One example of a phenomologically credible grand unified theory (GUT)
has the symmetry breaking
\bea
SO(10) & \pharrow{126} & SU(5) \times Z_2 \\ 
& \pharrow{45} & SU(3)_c \times SU(2)_L \times U(1)_Y \times Z_2 \\
& \pharrow{10} & SU(3)_c \times U(1)_Q \times Z_2 \ .
\eea
The discrete $Z_2$ symmetry allows the formation of topologically
stable cosmic strings. One possibility is an abelian string, in which
case
\be 
\phnb{126} = e^{in\th} \phi_0 f(r) \ ,
\ee
where $\phi_0$ is the usual VEV of \ph{126}, $f(0)=0$ and
$f(\infty)=1$. To give a vanishing covariant derivative at infinity
there is also a non-zero gauge field. The only stable abelian strings
have $|n|=1$, but higher winding number strings may have long
lifetimes. Nonabelian strings also form in this model, but they do not
have zero modes at high temperatures, and so will be ignored 
\cite{SO10strings}.

The string gauge field has a non-trivial effect on the electroweak Higgs
field, \ph{10}. Thus
\be
\phnb{10} = \left[H_0 e^{im\th} + H'_0 e^{-im\th}\right] h(r) \ ,
\ee
where $H_0$ and $H'_0$ are the usual VEVs of the components of
\ph{10}, $h(\infty)=1$, and $h(0)$ is small ($m=0$) or zero ($m \neq 0$). 
There is also a non-vanishing electroweak gauge field. $m$ is
determined by the GUT string, and is equal to the nearest integer to
$-n/5$, so the electroweak Higgs field does not wind around a
topologically stable abelian string.

Of the three Higgs fields, only \ph{126} and \ph{10} couple to
fermions. Only right-handed neutrinos couple to
$\phi_0$, while neutrinos of either helicity couple to $H'_0$.
The full mass matrix is a direct product of several
matrices. The neutrinos get their masses from
\be
M^{(\nu)} = \smat{\mgut f(r)e^{in\th}}
		{\mup h(r) e^{-im\th}}{\mup h(r) e^{-im\th}}{0} \ ,
\ee
where $\mup$ is the up quark mass and $\mgut \sim |\phi_0| \sim 10^{16}$GeV. 
In the notation of \bref{Lag} $\psi_2 = \nu_L$ and 
$\psi_1 = i\sigma^2\bar{\nu}_R^T$. The gauge eigenvalues of the two fields are
then $q_1 = n/2$ and $q_2 = -n/2 - m$. The other particles just couple to
$H_0$ or $H'_0$ \cite{SternYajnik}.

At high temperatures only \ph{126} is non-zero and $\mup=0$, so only
$\psi_1$ features in the zero mode analysis. Using
\bref{n2fer} gives the number of neutrino zero modes as
\be
\nl = \pos{n} \ , \ \nr = \pos{-n} \ .
\ee
At low temperatures $\mup \neq 0$. This time
\be
\nl = \pos{-2m} \ , \ \nr = \pos{2m} \ .
\ee
Since $|2m| < |n|$, some of the zero modes will be destroyed. For a
stable $n=1$ string all zero modes are destroyed. This is in agreement
with results obtained in \cite{SO10strings}. Thus, since higher $n$
strings almost certainly decay, there are zero modes before, but not
after the electroweak phase transition.  The neutral current in the
string disperses \cite{HillWidrow} and any vortons formed would
dissipate after about $10^{-10}$sec \cite{acd&wbp3}. Before the electroweak
phase transition from about $10^{10}$GeV -- $10^2$GeV the universe
would undergo a period of matter domination. Once the vortons
dissipate there would be some reheating of the universe. However the
electroweak interactions and physics below the phase transition would
be unaffected.

In a more arbitrary $U(1) \times $(Standard Model) theory, the ratio of
$m$ and $n$ could be greater than 1, in which case extra zero modes
would be created at the electroweak phase transition.

The destruction of zero modes at phase transitions will occur in a
wide range of models. In a theory with majorana mass terms, such as the
right handed neutrino term above, the first version of the index
theorem can be used \bref{allindex}. After a phase transition, if every fermion
couples to a non-winding Higgs field, such as the electroweak Higgs
field above, then all previously formed zero modes will be destroyed
(see remarks after \bref{allindex}).

\section{A Model with Persistent Zero Modes}
\label{Persistent}

In this section we consider a model in which zero modes survive a
subsequent symmetry breaking, despite coupling to a Higgs field with a
non-winding component. Consider fermions which have the usual Yukawa
couplings with two Higgs fields. Neglecting gauge fields, the form of
the Lagrangian for each pair of fermions is then
\be
\Lagf = - i\bar f_L \gamma^\mu\partial_\mu f_L 
	- i\bar f_R\gamma^\mu\partial_\mu f_R
	+ m(\bar f_L\phi^* f_R + \bar f_R\phi f_L) \ ,
\ee
where $\phi$ is a combination of the two Higgs fields, only one of
which gives rise to a string. This leads to the Dirac equations 
\be
 -i \gamma^\mu\partial_\mu f_L + m\phi^* f_R=0 \ ,
\ee
\be
 -i \gamma^\mu\partial_\mu f_R + m\phi f_L=0 \ .
\ee
Using the following forms for the left and right-handed spinors,
\be
f_L=\pmatrix{A\cr B\cr -A\cr -B\cr}\qquad f_R=\pmatrix{C\cr D\cr C\cr D\cr} \ ,
\ee
the equations of motion become

\be
\pmatrix{
w+k & i e^{-i\theta}(-\partial_r+{i\over r}\partial_\theta) & m\phi^* & 0 \cr
i e^{i\theta}(\partial_r +{i\over r}\partial_\theta) & -w+k  & 0 & m\phi^* \cr
-m\phi & 0 & -w+k  & i e^{-i\theta} (\partial_r-{i\over r}\partial_\theta) \cr
0 & -m\phi & i e^{i\theta}(-\partial_r- {i\over r}\partial_\theta) & w+k \cr
}
\pmatrix{ A\cr B\cr C\cr D\cr} = 0
\ee

We can now try the usual zero mode analysis \cite{Jackiw}; let $w=k$ and set
$A=D=0$. The equations of motion reduce to

\be 
-i e^{-i\theta}\left(\partial_r-{i\over r}\partial_\theta \right) B 
	+ m\phi^* C=0
\ee
and
\be
-m\phi B 
- i e^{i\theta}\left(\partial_r+ {i\over r}\partial_\theta\right) C=0 \ .
\ee

The Higgs field has two parts, a winding part from the string
and a constant part from a second symmetry breaking,
\be
\phi = f(r) e^{i\theta} + p \ .
\ee
Note that this form for $\phi$ cannot be written in the form used in (4)
and so the index theorem developed in section 2 does not apply in this
case.
Changing variables to $X$ and $Y$ with
\be
X = \int_0^r f(\rho) d\rho +pr\cos\theta +c \ ,
\ee
we have
\be
\left({\partial X\over \partial r}\pm 
		{i\over r}{\partial X\over \partial \theta}\right) 
	= f(r) + p\left[\cos\theta\pm {i\over r}(-r\sin\theta)\right] \ .
\ee
Taking $B$ and $C$ independent of $Y$ the equations of motion become
\be 
-i e^{-i\theta}[f(r)+p(\cos\theta+i\sin\theta)] \partial_X B + m\phi^* C=0
\ee
and
\be
-m\phi B -ie^{i\theta}[f(r)+p(\cos\theta-i\sin\theta)] \partial_X C=0 \ .
\ee
The Higgs fields cancel and we have
\be
(-iB)_{,X} +mC = 0\quad ,\quad m(-iB) + C_{,X}=0
\ee
which have a solution,
\be
\pmatrix{ (-iB)\cr C\cr}= e^{-mX}\pmatrix{1\cr 1\cr} \ .
\ee

We have explicitly constructed a zero mode after the second phase transition.
The $X$ coordinate is similar to the usual radial coordinate, but
'centres' on the effective zero of the resultant Higgs field, rather
than the core of the string.

\section{Index Theorems and Spectral Flow}
\label{Spectral}
\input{psfig}
\begin{figure}
\centerline{\psfig{file=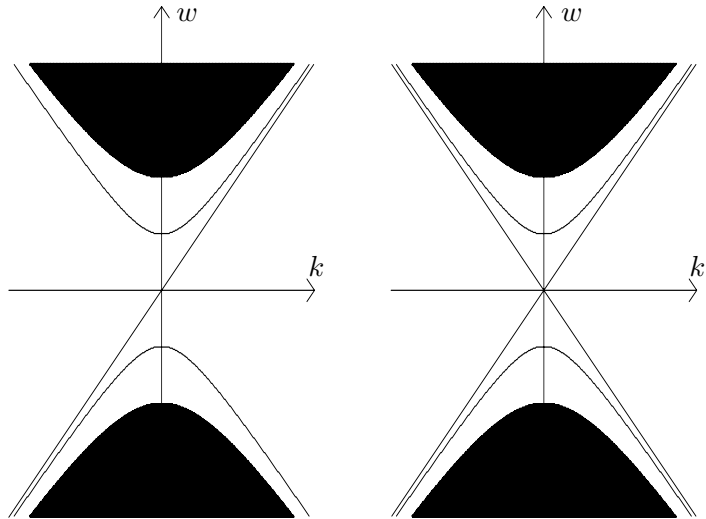,width=3.5in}}
\caption{The Dirac spectrum with a zero mode (left) and a very low
lying bound state (right). Both spectra also have a bound state and
continuum.}
\vskip -3.7 in 
\hskip 1.92 in $w$
\hskip 1.85 in $w$
\vskip 1.15 in 
\hskip 2.61 in $k$
\hskip 1.86 in $k$
\vskip 2.4 in
\bigskip
\label{specfig}
\end{figure}

We have shown that zero modes can acquire masses at subsequent phase
transitions. No matter how small this mass the spectrum of the Dirac
operator changes significantly. If we compare the Dirac spectrum 
with a zero mode 
and a low lying bound state with infinitesimal mass
(fig.~\ref{specfig}), we see that an arbitrarily
small perturbation to the zero mode introduces an entire new branch to
the spectrum. Any massive state gives a spectrum that is symmetric
about both the $w$ and $k$ axes, there is always a reference frame in
which the particle is at rest and others where it is moving up or
down the string. Conversely the zero mode, which is massless, can only
move in one direction along the string and its spectrum is asymmetric.
The transition from zero mode to low lying bound state causes drastic
changes in the spectrum and can be brought about by infinitesimal
changes in the value of one Higgs field. If we consider the species
with the zero mode alone, this infinite susceptibility to the
background fields appears unphysical. However, when we include the
massless neutrino in the $SO(10)$ model the spectral changes are less
worrying. For a small coupling between the two neutrinos, both the
before and after spectra have a continuum of massless or nearly
massless states. These states can be used to build the extra branch of
the perturbed zero mode spectrum, allowing small changes in the
overall spectrum for small changes in the background fields.

This observation leads us to conjecture that zero modes can be removed
only if they become mixed with other states. 

The coupling between the left and right handed neutrinos and the
electroweak Higgs field need not be artificially small, the small mass
of the light neutrino can be generated by the seesaw mechanism
\cite{seesaw}. This coupling is present prior to the electroweak phase
transition and allows transitions of the form $\nu_L+\bar\nu_R \to f \bar f$,
where $f$ is any light fermion from the standard model and the
intermediate state is an electroweak Higgs. Such interactions allow
zero modes on the string to scatter from massless neutrinos in the
surrounding plasma and provide a current damping mechanism that
affects current build up prior to the electroweak transition.

\section{Conclusions}

In this paper we have seen that the microphysics of cosmic strings
can be influenced by subsequent phase transitions. Fermion zero modes,
and consequently superconductivity, of the strings can be created
or destroyed by such phase transitions. In determining whether or
not a cosmic string is superconducting it is not enough to just consider
this at formation, but to follow the microphysics through the multiple 
phase transitions that the system undergoes. It is possible for
vortons formed at high energy to dissipate after a subsequent phase
transition if the relevant fermion zero mode does not survive the
phase transition, thus vortons bounds could be evaded. This is the
case with $\nu_R$ zero modes on the $SO(10)$ string \cite{SO10strings}. 
Prior to dissipation there could be a period of vorton
domination, after the phase transition the universe would reheat and
then evolve as normal.

To enable a systematic analysis of this effect we have derived a
generalised index theorem. Our index theorem is especially applicable 
to theories where the fermions acquire mass from more than one Higgs field.
We applied the index theorem and also considered
spectral flow. As a result we conjecture that zero modes are destroyed
when they mix with other fermions that acquire mass at a subsequent
phase transition from a non-winding Higgs field. 

\section{Acknowledgements}
This work was supported in part by PPARC, the EU under the HC programme
(CHRX-CT94-0423) and Trinity College.

\end{document}